# Dissipative soliton comb

Evgeniy V. Podivilov[1,2], Denis S. Kharenko[1,2], Anastasia E. Bednyakova[2,3], Mikhail P. Fedoruk[2,3], Sergey A. Babin[1,2*]

[1] Institute of Automation and Electrometry, SB RAS, Novosibirsk 630090, Russia
[2] Novosibirsk State University, Novosibirsk 630090, Russia
[3] Institute of Computational Technologies, SB RAS, Novosibirsk 630090, Russia
*Corresponding author: babin@iae.nsk.su

**Dissipative solitons are stable localized coherent structures with linear frequency chirp generated in normal-dispersion mode-locked lasers. The soliton energy in fiber lasers is limited by the Raman effect, but implementation of intracavity feedback for the Stokes wave enables synchronous generation of a coherent Raman dissipative soliton. Here we demonstrate a new approach for generating chirped pulses at new wavelengths by mixing in a highly-nonlinear fiber of two frequency-shifted dissipative solitons, as well as cascaded generation of their clones forming a "dissipative soliton comb" in the frequency domain. We observed up to eight equidistant components in a 400-nm interval demonstrating compressibility from ~10 ps to ~300 fs. This approach, being different from traditional frequency combs, can inspire new developments in fundamental science and applications.**

It is well known that the phase synchronization of laser modes (so-called mode locking) forms a pulse train with a period equal to the inverse mode spacing and duration equal to the inverse gain bandwidth [1]. In the case of a broadband gain medium, the mode-locked laser can generate ultrashort pulses while its broad spectrum consisting of equidistant frequencies represents so-called "frequency comb" [2,3]. In Ti:sapphire lasers, the comb width can reach one octave [4]. Such a broad coherent spectrum called "supercontinuum" (SC) has opened a new area of optics – frequency metrology [2].

A coherent SC can also be generated outside the laser cavity, e.g. by launching the mode-locked laser pulse (that can form a soliton in anomalous-dispersion regime) into a highly nonlinear medium, either bulk crystals/silica [5] or silica-based fiber waveguides, in particular, tapered ones or photonic-crystal fibers (PCF) characterized by high nonlinearity and controllable dispersion [6]. The effects induced by the Kerr nonlinearity, such as self-phase modulation (SPM) or cross-phase modulation (XPM), broaden the spectrum, while the increasing group velocity dispersion limits the SC bandwidth [6–9]. The Kerr nonlinearity effects for multi-frequency radiation can be treated in the frequency domain as cascaded four-wave mixing (FWM) between corresponding modes/frequencies [8,9] which results in generation of additional components, thus forming a coherent comb. Kerr frequency combs can be also generated via FWM parametric oscillation in microresonators [10], which are especially attractive for mid-IR range [11,12]. This technique also enables mode locking and soliton formation [12-14]. Simultaneous second- and third- harmonic conversion of a Kerr comb within the same microresonator leads to generation of comb-like visible lines [15].

Dissipative solitons (DS) generated in normal-dispersion mode-locked lasers [16,17] are characterized by linear frequency modulation (chirp) along the pulse. As shown recently, Stokes-shifted Raman dissipative solitons (RDS) can be synchronously generated in the same laser cavity [18,19]. In this paper, we study the mixing between these nearly identical coherent chirped pulses with different carrier frequencies in a highly nonlinear medium (PCF as an example). It appears that multiple equidistant chirped pulse lines are generated at harmonics of the DS-RDS frequency difference both in long- and short-wavelength domains thus forming a "dissipative soliton comb". The theoretical background, proof-of-principle experiments and potential applications of this new approach are described below.

Let us treat two (±) coherent pulses with amplitude $A_\pm(t)$ and carrier frequency $\omega_\pm$. The amplitude of the combined field is

$$E(t) = A_+(t)\exp(i\omega_+ t) + A_-(t)\exp(i\omega_- t) \quad (1)$$

Assume that the frequency difference $\Delta_{st} = \omega_+ - \omega_-$ is much larger than the pulse spectral width $\Delta_\pm$. Launching the pulses into a PCF having zero dispersion near $\omega_\pm$ will result in generation of new spectral components at $\omega_{1\pm}=\omega_{\pm}\pm\Delta_{st}$ via the FWM process. Their amplitudes $A_{1\pm}$ grow linearly with distance $z=ct/n$:

$$A_{1\pm}(t)=i\gamma z A_{\pm}^2(t) A_{\mp}^*, \quad (2)$$

where $\gamma$ is the Kerr nonlinearity coefficient. Similar four-wave model was used e.g. for modeling of micro-resonators with CW bi-chromatic pumping [20].

In the first perturbation order, the spectral satellite power $P_{1\pm}=A_{1\pm}^2$ is lower than the pump power $P_{\pm}$ by factor $\rho=\gamma^2 z^2 P_+ P_- <<1$. Increasing satellite amplitude $A_{1\pm}$ initiates FWM between the satellites and pump pulses thus resulting in generation of higher-order components $\omega_{\pm}\pm 2\Delta_{st}$ with the power lower by $\rho^2/2$ times as compared to the pump power. Herewith, propagation of the pump pulses is accompanied by the variation of their phase via the SPM and XPM processes as:

$$A_{\pm}^{out} = A_{\pm}(t)\exp\left[i\gamma z\left(P_{\pm}(t)+2P_{\mp}(t)\right)\right]$$

For transform limited pulses (e.g. solitons) with temporal and spectral half-widths $T$ and $\Delta$ ($T\cdot\Delta\sim 1$), the relative spectral broadening is $\delta\Delta/\Delta\sim 3\rho$. The corresponding qualitative picture in the time-frequency plane is shown in Fig.1a. The process of spectral broadening can be represented as FWM between internal frequency components (modes) with spacing $\delta$ resulting in generation of new components separated by $\delta$. When we mix two mutually coherent solitons with frequency separation $\Delta_{st}$, which overlap in time, additional components separated by $\Delta_{st}$ appear after passing PCF. The processes of new components generation ($\sim\Delta_{st}$) and pulse broadening ($\sim\delta$) occur simultaneously (Fig.1a), finally leading to the formation of a continuously broadened frequency comb.

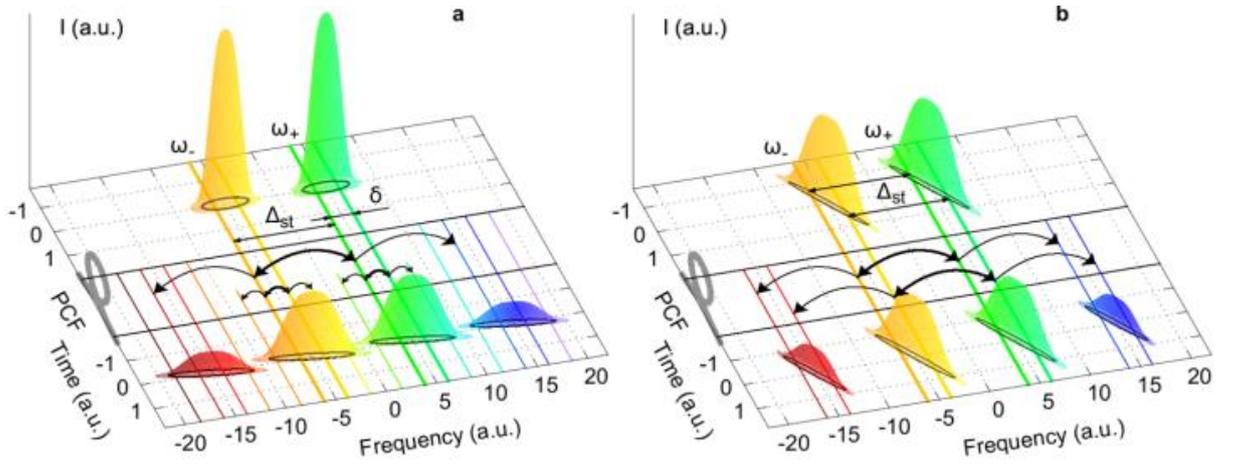

**Fig 1. Nonlinear mixing in a PCF** of two solitons (a) and two equally chirped dissipative solitons (b) with frequency separation $\Delta_{st}$, each of which consisting of laser modes with separation $\delta$.

For chirped dissipative solitons of the same spectral width $\Delta$, the picture is principally different. The DS features an increased time-bandwidth product, $T \cdot \Delta \sim f \gg 1$, as the frequency chirp stretches the pulse along the diagonal of the $\Delta$-$T$ square (Fig.1b). If two input pulses with carrier frequencies $\omega_{\pm}$ have a similar chirp parameter $f$, the frequency spacing $\Delta_{st}=\omega_+ -\omega_-$ remains unchanged from the front to the trailing edge of the pulses, when they propagate in a dispersion-free PCF. Therefore, satellites at $\omega_{\pm} \pm \Delta_{st}$ generated in the FWM process should be also chirped. Let us treat two Gaussian input pulses $A_{\pm}(t) = A_{\pm}\exp(-t^2(1+if_{\pm})/2T_{\pm}^2)$ with a corresponding half-width $T_{\pm}$ and chirp parameter $f_{\pm} \gg 1$. Their spectrum is also Gaussian: $I_{\pm}(\omega) \propto \exp(-(\omega-\omega_{\pm})^2/\Delta_{\pm}^2)$, where $\Delta_{\pm} = (1+f_{\pm}^2/T_{\pm}^2)^{1/2} \approx f_{\pm}/T_{\pm}$ is the spectral half-width. According to (2), satellites born in the FWM process have duration $T_{1\pm} = (T_{\pm}^{-2} + 2T_{\mp}^{-2})^{-1/2}$, chirp parameter $f_{1\pm} = (2f_{\pm}T_{\pm}^{-2} - f_{\mp}T_{\mp}^{-2})T_{1\pm}^2$ and spectral width $\Delta_{1\pm} \approx f_{1\pm}/T_{1\pm}$. In the case of two identical pump pulses, the satellites temporal/spectral widths are reduced by factor $\sqrt{3}$ in the first perturbation order. At the same time, the SPM/XPM-induced spectral broadening $\delta\Delta \sim 3\rho/T$ is small relative to the linewidth, $\delta\Delta/\Delta \sim 3\rho/f$, as $f \gg 1$. In terms of the FWM process, it means that the mixing between intra-pulse components $\delta$ is suppressed because they do not overlap in time (see Fig.1b).

With increasing length and/or power, parameter $\rho$ grows up to unity and multiple lines in the low-dispersion spectral window should be generated with the same amplitude and width, i.e. a comb of similar chirped pulses will arise. However, representation of nonlinear process as SPM, XPM, FWM is not possible at high powers, so a more general description is necessary in this case. It was developed for identical input pulses $A_+(t) = A_-(t) = A(t)$. Neglecting dispersion, the combined pulse envelope $B(t) = 2A(t)\cos(\Delta_{st}/2)$ evolves as:

$$\frac{\partial B}{\partial z} - \frac{1}{c}\frac{\partial B}{\partial t} = i\gamma |B|^2 B. \quad (3)$$

The solution at the output of PCF with length L is

$$B^{out}(t) = B(t)\exp\left[2i\gamma LI(t)\left(1+\cos(\Delta_{st}t)\right)\right]$$

$$= Ae^{2\gamma LI(t)} \sum_{n=-\infty}^{\infty} e^{in\Delta_{st}t} i^n \left[J_n(2\gamma LI(t)) + iJ_{n+1}(2\gamma LI(t))\right], \quad (4)$$

where $I(t) = |A(t)|^2$ is the intensity of each input pulse. Eq. (4) shows that the output signal represents a "comb" of equidistant spectral components separated by $\Delta_{st}$. A satellite with number $n$ takes the form of the input signal $A(t)$ multiplied by the sum of two Bessel functions $J_n$, $J_{n+1}$. For the input pulses in the form of a highly chirped dissipative soliton (solution of the Ginzburg-Landau equation) [16,21]: $A(t) = A_0 \cosh^{(if-1)}(t/T)$ with half-width $T$ and chirp parameter $f$, we can obtain Fourier transform expressed via the $\beta$ function. However, in the high-chirp limit ($f \gg 1$) an analytical expression can be obtained [22]. The output spectrum of the $n$-th satellite takes the following form:

$$P_n(\Omega) = P\left(J_n^2\left(2\gamma LI_0\left(1-\frac{\Omega^2}{\Delta^2}\right)\right) + J_{n+1}^2\left(2\gamma LI_0\left(1-\frac{\Omega^2}{\Delta^2}\right)\right)\right), \quad (5)$$

where $\Omega$ is the frequency detuning, and $\Delta = f/T$ is the pulse spectrum half-width. For high-intensity input pulses, the amplitudes of far (large $n$) satellites increase, while the spectral profiles of near satellites become disturbed.

We can also account for the dispersion and Raman effects in numerical simulations. The signal evolution inside the nonlinear fiber is described by the generalized nonlinear Schrödinger equation (NLSE):

$$\frac{\partial A}{\partial z} = \sum_{i=2}^{7} \frac{i^{k+1}}{k!}\beta_k \frac{\partial^k A}{\partial t^k} + i\gamma\left(A(z,t)\int_0^\infty R(t')|A(z,t-t')|^2 dt'\right), \quad (6)$$

where $A(z,t)$ is the electric field envelope, $\beta_k$ are the dispersion coefficients at the central frequency $\omega_0$, $\gamma$ is the Kerr nonlinearity coefficient. The response function $R(t)=(1-f_R)\delta(t)+f_R h_R(t)$ includes both instantaneous electronic and delayed Raman contributions, where $f_R$ represents the fractional contribution of the delayed Raman response to the instantaneous nonlinear polarization [21].

The output pulses were modeled analytically and numerically for the SC-5.5-1040 fiber (NKT Photonics Ltd.) with the parameters presented in Table 1, including Kerr nonlinearity ($\gamma$) and Raman gain ($g_R$) coefficients, as well as dispersion coefficients associated with the Taylor series expansion of the propagation constant $\beta(\omega)$ near zero dispersion wavelength $\lambda_0$. The full dispersion curve is shown in Fig. 2 together with the results of calculations.

Table 1. Parameters of PCF SC-5.5-1040

| Par | Value | Unit | Par | Value | Unit |
|---|---|---|---|---|---|
| $\lambda_0$ | 1040 | nm | $\beta_3$ | $7.3365\times10^{-2}$ | $ps^3/km$ |
| $\omega_0$ | $2\pi c/\lambda_0$ | rad | $\beta_4$ | $-1.0891\times10^{-4}$ | $ps^4/km$ |
| $\gamma$ | 11 | $W^{-1}km^{-1}$ | $\beta_5$ | $1.4161\times10^{-7}$ | $ps^5/km$ |
| $g_R$ | 5.3 | $W^{-1}km^{-1}$ | $\beta_6$ | $-1.0986\times10^{-9}$ | $ps^6/km$ |
| $\beta_2$ | $3.1737\times10^{-2}$ | $ps^2/km$ | $\beta_7$ | $7.5900\times10^{-12}$ | $ps^7/km$ |

The light blue line in Fig. 2(a) depicts the Fourier transform of the analytical solution (4), where A(t) has the form of a chirped DS, calculated numerically according to the model of the DS-RDS generator described in [18,19]. The color line shows the output spectrum, calculated with Eq. (5), the corresponding parameters of the input pulse (hyperbolic cosine) are $P_0 = |A_0|^2 = 180$ W, $T = 15$ ps, and $f = 175$. The dashed black line shows the numerical solution of the generalized NLSE (Eq. 6), obtained with the same initial condition as the analytical solution at zero dispersion and

Raman gain. Fig. 2(a) shows that the analytics work quite well in the dispersion-free Raman-free case.

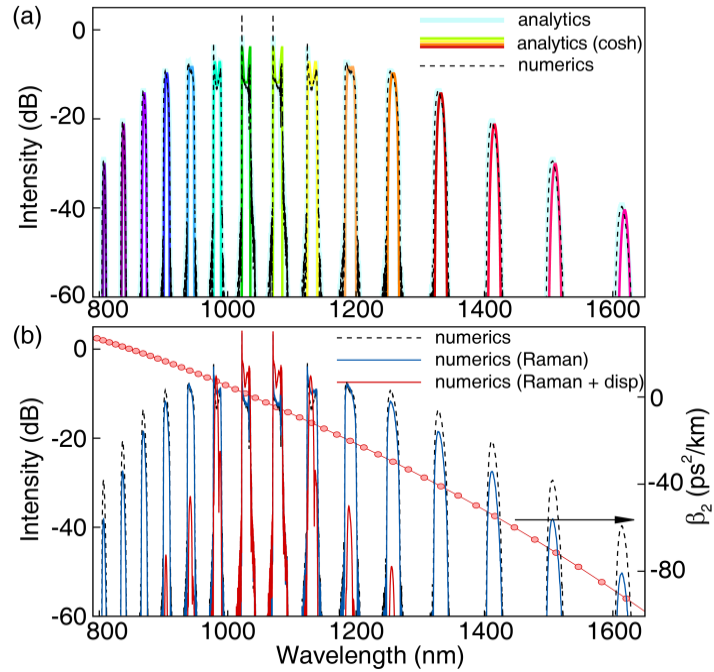

**Fig. 2: Parameters and results of calculations for PCF SC-5.5-1040:** output spectra calculated with analytical expressions (3)-(4) (blue line), or (5) (color line), and numerically (solution of the NLSE (6)) with neglected dispersion and Raman effects (a) and with account for Raman and dispersion (coefficient $\beta_2$ is shown by circles) effects (b)

The blue line in Fig. 2(b) shows the solution of the generalized NLSE, where the Raman gain in PCF is taken into account. The red line refers to the case with both Raman gain and dispersion taken into account. One can see that dispersion could significantly decrease the number of spectral satellites (clones), generated at the DS-RDS propagation in a nonlinear fiber, while the Raman effect is not so strong.

To perform a proof-of-principle experiment the following setup was assembled, see Fig. 3. We test a 1-meter PCF (SC-5.5-1040) with two coherent chirped pulses with wavelengths around 1040 nm generated in the DS-RDS generator. It consists of a 30-m long PM fiber (PMF), isolator (ISO), polarization controller (PC) and non-PM Yb-doped fiber amplifier (YDFA) enabling generation of a highly-chirped DS centered at 1025 nm. Inserting a delay line (DL) with a spectrally-selective PWDM splitter and a fiber coupler with a feedback coefficient R=10$^{-4}$ at the Stokes wavelength provides synchronous generation of a Stokes-shifted Raman DS at 1070 nm, according to the technique proposed in [18,19]. The generated RDS leaves the cavity through the 99% port of the coupler, while the DS exits through the orthogonal-polarization port of the PBS. The DS and RDS generated in a common cavity are shown to be mutually coherent [19], that is a question in case of external-cavity RDS generation demonstrated recently for fiber and diamond Raman lasers [23,24].

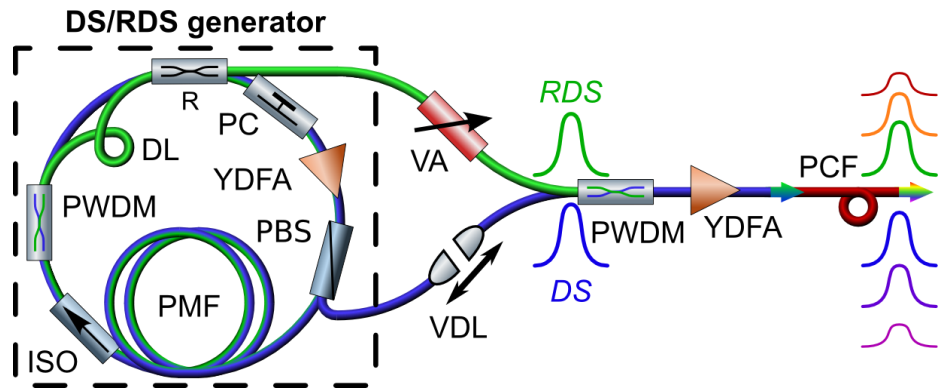

**Fig 3. Experimental setup:** The DS/RDS generator has two fiber outputs for the DS and RDS, which are combined by polarization wavelength division multiplexer (PWDM) after temporal and amplitude equalization by a variable delay line (VDL) and a variable attenuator (VA). After amplification by Yb-doped fiber (YDFA) they propagate together in the PCF thus generating their clones at new wavelengths both in short (anti-Stokes) and long (Stokes) wavelength domains. All components are PM, except for the intra-cavity YDFA with pigtails and the PCF.

After matching the DS and RDS by a variable delay line (VDL) and equalization by a variable attenuator (VA), they are combined with proper polarization at the external coupler (PWDM) and are launched into the PCF via loss-compensating YDFA. The mixed pulses have duration of about 20 ps and total energy up to 16 nJ. At the PCF output, new equidistant spectral components are observed, as predicted by the simulation under conditions of the experiment (see Fig. 4). The initial field at the PCF input, shown in the inset of Fig. 4(a), is taken from the DS/RDS fiber laser model [18,19] for the pulse energy equally divided between the DS and RDS

(8 nJ each). The output spectrum presented in Fig. 4(a) is calculated numerically for the experimental conditions with account for the dispersion and Raman effects in the PCF. New spectral components appear at >1100 nm and <1000 nm, the most eminent of which are the first Stokes and anti-Stokes pulses at 1110 and 980 nm with 1.6 and 1.7 nJ energies, respectively. At that, the DS and RDS pump pulses are noticibly depleted at the output due to the energy transfer to the Stokes and anti-Stokes components.

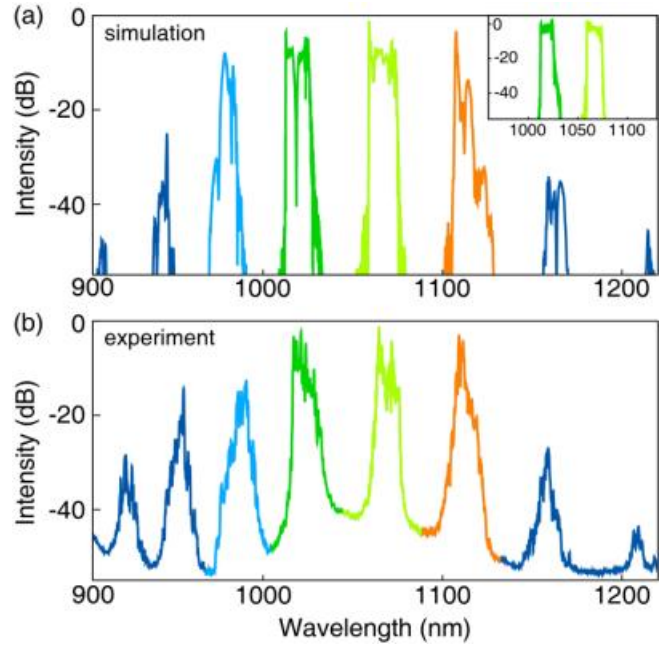

**Fig 4. Output spectra** in simuation (a) and experiment (b). Inset: spectral shape of the DS and RDS at the input of the nonlinear fiber.

The experimental spectra measured by an optical spectrum analyzer (Yokogawa 6370) (see Fig. 4(b)) as well as the estimated total energy of satellites (2-4 nJ depending on adjustments at the input) are quite similar to those in calculations. We also measured temporal characteristics of each satellite by the FROG (Mesaphotonics Ltd.) and intreferometric ACF (Avesta Ltd AA-20DD autocorrelator) techniques before and after a grating compressor. The FROG trace was retrieved by the VideoFROG Scan software. The pulse train was controlled by a 1 GHz photodiode (New Focus 1601FS-AC). The results are shown in Fig. 5(a,b) for the first anti-Stokes (~990 nm) and Stokes (~1110 nm) satellites demonstrating nearly a linear chirp at a duration of ~10 ps compressible to 300-800 fs. The satellites can be treated as clones of the input dissipative solitons, whose number is limited by dispersion. Note that the mixing of chirped pulses in a long PCF with high dispersion is defined by the phase matching condition enabling generation of only one narrow (~1nm) Stokes component tunable within a ~20-nm range by varying the delay between the input pulses [25]. A short PCF placed in a cavity of optical parametric amplifier or oscillator also generates only one component (Stokes or anti-Stokes), see e.g. [26] and citation therein.

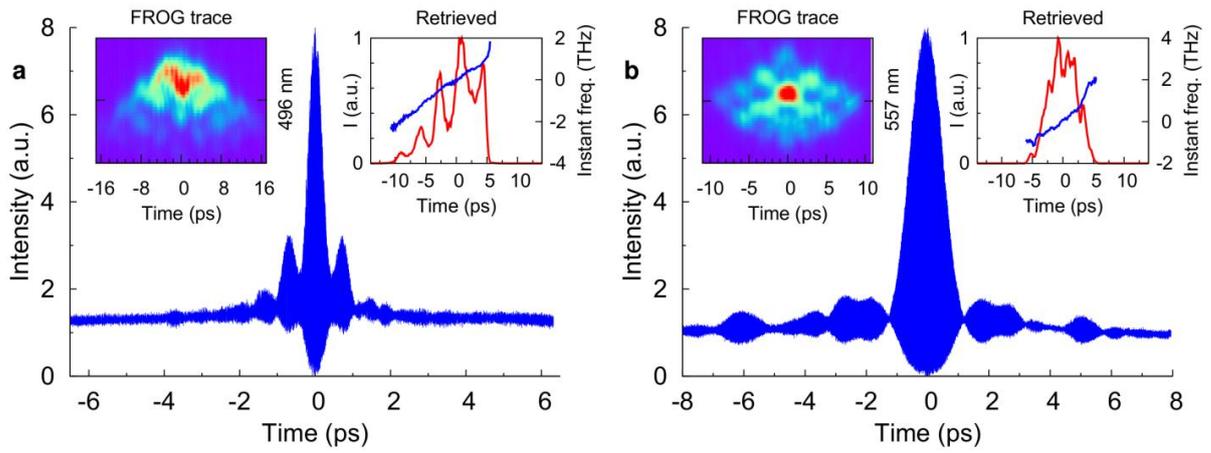

**Fig 5. Temporal characteristics of the generated anti-Stokes (a) and Stokes (b) pulses**: Interferometric ACF for a compressed pulse, FROG trace (left inset) and retrieved pulse amplitude and instantaneous frequency (right inset).

Thus, we have proposed and demonstrated theoretically and experimentally a new approach for generating chirped pulses at new wavelengths by mixing of two frequency-separated dissipative solitons in a highly-nonlinear fiber, as well as cascaded generation of their clones. Multiple clones are separated in frequency by the same interval and have similar characteristics (spectrum, duration, and chirp) as input DSs. The energy of an individual component reaches 4 nJ for ~8 nJ pump pulses and the compressed duration approaches ~300 fs. Up to 8 output components in a 400-nm spectral interval are observed in the experiment with a 1-m PCF. Further increase of input pulse energies and/or using a PCF with a broader low-dispersion window may result in an octave-spanning dissipative soliton comb. The comb period may be adjusted by means of the frequency difference variation between the DS and RDS or doubling it in the second-order RDS scheme [27]. This approach can be also transferred to other spectral regions by using other types of fiber lasers for generation of DS/RDS pump pulses, e.g. Er (1.55 µm) or Tm, Ho (2 µm) [28], as well as to microresonators, which also exhibit different Raman effects such as Raman self-frequency shift of Kerr solitons [29], Raman frequency comb formation [30], and synchronous Stokes soliton generation in presence of the main soliton [31]. Together with the input to fundamental science, the proposed approach offers quite new opportunities for applications. A proper phase correction and coherent combination of the dissipative soliton comb components may be used for generation of high-energy few-cycle pulses and/or for arbitrary waveform synthesis, analogues to that with the conventional frequency comb (coherent supercontinuum) [32] or with several synchronized mode-locked or CW lasers operating in different spectral ranges [33,34]. Compared to independent sources, the approach of the dissipative soliton comb is intrinsically stable and much simpler in realization. Such coherent dissipative soliton comb centered at 1.55 micron can be implemented in ultra-broadband transmission lines with new coherent modulation/demodulation formats [35]. Other applications can also benefit from this approach, including frequency comb spectroscopy [36], coherent biomedical imaging and microscopy [26], mid-IR and THz generation [37] and others.

**Funding.** We acknowledge support of Russian Science Foundation (grant 14-22-00118, work of D. S. Kh., E.V.P. and S. A. B; and grant 14-21-00110, work of A. E. B and M. P. F.).